\documentclass[aps,prl,twocolumn,groupedaddress]{revtex4}

\newcommand{\ket}[1]{\ensuremath{\left| {#1} \right>}}
\newcommand{\bra}[1]{\ensuremath{\left< {#1} \right|}}
\newcommand{\ave}[1]{\ensuremath{\left< {#1} \right>}}

\newcommand{\upup}{\uparrow\uparrow}
\newcommand{\up}{\uparrow}
\newcommand{\down}{\downarrow}
\newcommand{\updown}{\uparrow\downarrow}
\newcommand{\downup}{\downarrow\uparrow}
\newcommand{\downdown}{\downarrow\downarrow}
\newcommand{\uu}{\ket{\upup}}
\newcommand{\ud}{\ket{\updown}}
\newcommand{\du}{\ket{\downup}}
\newcommand{\dd}{\ket{\downdown}}
\newcommand{\Beplus}{\ensuremath{{^9}{\rm Be}^{+} \,}}
\newcommand{\Mgplus}{\ensuremath{{^{24}}{\rm Mg}^{+} \,}}
\newcommand{\Mgplusbf}{\ensuremath{\mathbf{^{24}}\mathbf{Mg}\boldsymbol{^+}
\,}}

\RequirePackage{bm}
\usepackage{amsmath, amssymb, graphics}
\usepackage{graphicx}
\usepackage{lscape}
\usepackage{longtable} 	% for the longtable environment
\usepackage{array}

\begin{document}

%\preprint{}

\title{Entangled Mechanical Oscillators}


\author{J. D. Jost}
\email[To whom correspondence should be addressed: ]{john.d.jost@gmail.com}
%\thanks{}

\author{J. P. Home}

\author{J. M. Amini}

\author{D. Hanneke}

\author{R. Ozeri}
\altaffiliation{\emph{Present Address} Department of Physics of Complex Systems,
       Weizmann Institute of Science, Rehovot, 76100, Israel}
\author{C. Langer}
\altaffiliation{\emph{Present Address} Lockheed Martin, Denver, CO, 80127, U.S.A.} 
\author{J. J. Bollinger}
\author{D. Leibfried}
\author{D. J. Wineland}
\affiliation{Time and Frequency Division,
       NIST, Boulder, CO 80305, U.S.A.}


\date{\today}

\begin{abstract}

Hallmarks of quantum mechanics include superposition and entanglement. In the context of large complex systems, these features should lead to situations like Schr\"odinger's cat\cite{Schrodinger1935}, which exists in a superposition of alive and dead states entangled with a radioactive nucleus. Such situations are not observed in nature. This may simply be due to our inability to sufficiently isolate the system of interest from the surrounding environment\cite{08Ball, 08Schlosshauer} -- a technical limitation. Another possibility is some as-of-yet undiscovered mechanism that prevents the formation of macroscopic entangled states\cite{03Bassi}. Such a limitation might depend on the number of elementary constituents in the system\cite{02Leggett} or on the types of degrees of freedom that are entangled. Tests of the latter possibility have been made with photons, atoms, and condensed matter devices\cite{Zeilinger08,08NatureInsight}. One system ubiquitous to nature where entanglement has not been previously demonstrated is distinct mechanical oscillators. Here we demonstrate deterministic entanglement of separated mechanical oscillators, consisting of the vibrational states of two pairs of atomic ions held in different locations. We also demonstrate entanglement of the internal states of an atomic ion with a distant mechanical oscillator. These results show quantum entanglement in a degree of freedom that pervades the classical world. Such experiments may provide pathways towards generation of entangled states of larger scale mechanical oscillators\cite{Mancini02, 05Schwab, 08Kippenburg}, and offer possibilities for testing non-locality with mesoscopic systems\cite{Haroche05}. In addition, the control developed in these experiments is an important ingredient to scale up quantum information processing based on trapped atomic ions\cite{bible, Cirac00, 02Kielpinski}.
\end{abstract}

%\keywords{}

\maketitle


%%%%%%%%%%%%%%%%%%%%%%%%%%%%%%%%%%%%%%%%%%%%%%%%%%%%%%%%%%%%%%%%%%%%%%%%%%%%%%%%%%%%%%%%%%%%%%%%%%
%\section*{Introduction}%%%%%%%%%%%%%%%%%%%%%%%%%%%%%%%%%%%%%%%%%%%%%%%%%%%%%%%%%%%%%%%%%%%%%%%%%%%%
%%%%%%%%%%%%%%%%%%%%%%%%%%%%%%%%%%%%%%%%%%%%%%%%%%%%%%%%%%%%%%%%%%%%%%%%%%%%%%%%%%%%%%%%%%%%%%%%%%%%

Mechanical oscillators pervade nature; examples include the vibrations of violin strings, the oscillations of quartz crystals used in clocks, and the vibrations of atoms in a molecule. Independent of the size of the system, each mode of vibration can be described by the same equations that describe the oscillations of a mass attached to a fixed object by a spring. For very low energy oscillations, quantum mechanics is needed for a correct description: the energy is quantized and the motion can be described generally by superpositions of wavefunctions corresponding to each quantum level. Coherent states behave very much like classical oscillators, while other states have properties with distinctly non-classical features\cite{Schleich01}. Quantum mechanics also permits superposition states of multiple systems called entangled states, where the measured properties of the systems are correlated in ways that defy our every-day experience\cite{Aspect02,Zeilinger08, Matsukevich08,08NatureInsight,08Pan}. When extended to macroscopic scales, situations akin to Schr\"odinger's cat should appear. Our inability to produce such macroscopic entanglement may be just a question of technical difficulty.  However, there might be a more fundamental cause, such as the inability to entangle certain types of degrees of freedom.

To explore the latter territory in a new regime, we demonstrate entanglement of two separated mechanical oscillators.  Here each oscillator is comprised of a pair of ions - one \Beplus and one \Mgplus - confined in a potential well. In the context of the experiment described below, each pair behaves like two masses connected by a spring of length $\sim$ 4 $\mu$m, undergoing vibrational motion (Fig. 1a). The two pairs are separated by 0.24 mm such that the coupling between them can be neglected. To create the entangled state of the oscillators, we start with all four ions in one location and entangle the internal states of the two \Beplus ions\cite{didi_gate}. We then separate the four ions into two pairs, each containing one of the entangled \Beplus ions. Finally, we transfer the entanglement from the \Beplus ions' internal states to the motion of the separated ion pairs, creating the desired motional entanglement.

\begin{figure}
  \includegraphics[width=\columnwidth]{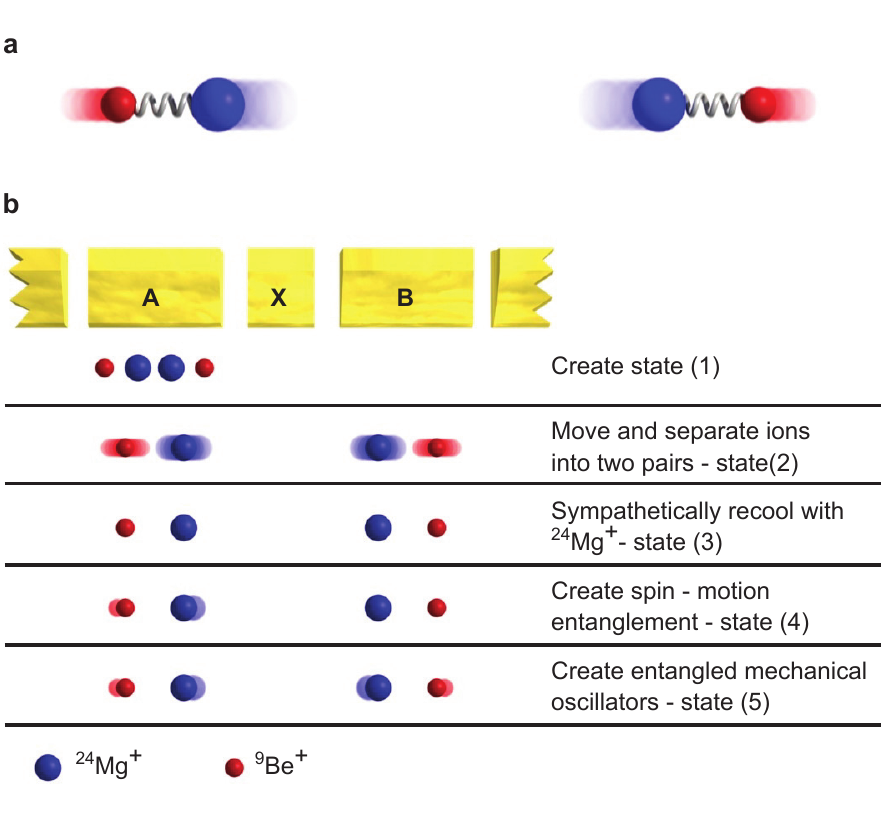}\\
    \caption{\textbf{Creation of entangled mechanical oscillators}. \textbf{a}, Simplified depiction of the two mechanical oscillators indicating motion in the stretch mode of each \Beplus -- \Mgplus ion pair, held in separate locations (not to scale). The pairs -- spaced by $\simeq$ 0.24 mm -- each behave as two masses spaced by $\sim$ 4 $\mu$m, connected by springs. \textbf{b}, Schematic showing the ions' positions with respect to the ion trap electrodes ($A$, $X$, and $B$) and the quantum states at key points in the experiments (not to scale). The entangled spin state (1) of the \Beplus ions is created with all the ions in a single well. All ions are moved adjacent to electrode $X$, where they are separated into two pairs, which then are distributed to potential wells located adjacent to electrodes $A$ and $B$ (state (\ref{state2})). Laser cooling of the \Mgplus ions removes motional excitation incurred during the separation process, reducing the motion to near the ground state (state (\ref{state3})). A laser induced stimulated-Raman sideband pulse on the \Beplus ion in well $A$ approximately creates state (\ref{spinmo}), where the motion in well $A$ is entangled with the \Beplus spin in well $B$. A subsequent pulse on the \Beplus ion in well $B$ approximately creates state (\ref{moent}) at which point the two mechanical oscillators are entangled.}
\end{figure}

%%%%%%%%%%%%%%%%%%%%%%%%%%%%%%%%%%%%%%%%%%%%%%%%%%%%%%%%%%%%%%%%%%%%%%%%%%%%%%%%%%%%%%%%%%%%%%%%%
%\section*{Initial State Prep}
%%%%%%%%%%%%%%%%%%%%%%%%%%%%%%%%%%%%%%%%%%%%%%%%%%%%%%%%%%%%%%%%%%%%%%%%%%%%%%%%%%%%%%%%%%%%%%%%%
Initially, all the ions are held in a single potential well of a multi-zone linear Paul trap\cite{rowe02,barrett04}.  The potential well is  configured to locate the ions along a line corresponding to the axis of weakest confinement, which we call the axial direction. We will be concerned only with motional modes along this axis. While applying continuous laser cooling, we initialize the ions in a particular order, \Beplus -- \Mgplus -- \Mgplus -- \Beplus, by first increasing the axial confinement until no linear arrangement is stable. The axial potential is independent of ion mass while the radial potential strength scales inversely with the mass\cite{bible}, thus there exist axial potentials where the heavier \Mgplus ions are displaced from the axis and must reside between the \Beplus ions. We then relax the axial confinement giving the desired order\cite{privatetill}.

Lasers provide control of the ions' motion and internal states through laser cooling and stimulated-Raman carrier or sideband transitions\cite{king98,bible,08NatureInsight}. Using Doppler cooling on the \Beplus and \Mgplus ions, followed by sideband cooling on the \Beplus ions, we prepare the motion of each of the four axial normal modes (see Methods) to an average motional occupation of $\ave{n} \le $ 0.17. By applying a magnetic field of 0.012 T, we spectrally isolate two internal (hyperfine) states in each \Beplus ion, which we call ``spin" states, and label $\ket{\up} \equiv \ket{F = 2, m_{F} = 2}$ and $\ket{\down} \equiv \ket{F = 2, m_{F} = 1}$. These states are split by 102 MHz.  Using a geometric phase gate\cite{didi_gate} and spin rotations, we create the decoherence-free-subspace entangled state
\begin{equation}
\ket{\Psi^{+}} =\textstyle{\frac{1}{\sqrt{2}}} \Bigl[\ket{\up\down} + \ket{\down\up}\Bigr]
\end{equation}
of the two \Beplus ions. This state is resistant to decoherence from spatially uniform magnetic field noise\cite{kielpinski01}.

%%%%%%%%%%%%%%%%%%%%%%%%%%%%%%%%%%%%%%%%%%%%%%%%%%%%%%%%%%%%%%%%%%%%%%%%%%%%%%%%%%%%%%%%%%%%%%%%%%%%
%\section*{Separation and Recooling}
%%%%%%%%%%%%%%%%%%%%%%%%%%%%%%%%%%%%%%%%%%%%%%%%%%%%%%%%%%%%%%%%%%%%%%%%%%%%%%%%%%%%%%%%%%%%%%%%%%%%

Time-varying axial potentials move and separate\cite{rowe02,barrett04} the four ions into two \Beplus -- \Mgplus pairs in different wells, which are spaced by $\sim$ 0.24 mm (see Fig. 1b). Each pair of ions has two axial normal modes: the ``stretch" mode (frequency $\simeq$ 4.9 MHz) in which the two ions oscillate out-of-phase and the ``common" mode ($\simeq$ 2.3 MHz) where they oscillate in-phase. The experiment involves the ground $\ket{n=0}_{j}$ and first excited $\ket{n=1}_{j}$ states of the stretch modes, where $j \in \{A, B\}$ refers to the well. In general, the separation process excites the motional modes into unknown states. The wavefunction of the \Beplus spin states after separation is
\begin{equation}
\label{state2}\textstyle{\frac{1}{\sqrt{2}}} \Bigl[\ket{\up}_{A}\ket{\down}_{B} + e^{i\xi(t)}\ket{\down}_{A}\ket{\up}_{B}\Bigr],
\end{equation}
where $\xi(t)$ is a phase that accumulates through the course of the experiment due to a small difference in magnetic field between wells $A$ and $B$.

To create the motional entangled state we first prepare the stretch modes close to $\ket{0}_{A}\ket{0}_{B}$. For this, Doppler and sideband laser cooling on the \Mgplus ions in both wells sympathetically cools\cite{barrett03} the \Beplus ions and prepares the stretch modes to mean occupation numbers of $\ave{n_{A}} =$ 0.06(2) and $\ave{n_{B}} =$ 0.02(2). We also cool the common mode in each well to $\ave{n} \leq$ 0.13. The cooling does not affect the spin states of the \Beplus ions\cite{barrett03}, thereby approximating the state
\begin{equation}
\label{state3}\textstyle{\frac{1}{\sqrt{2}}} \Bigl[\ket{\up}_{A}\ket{\down}_{B} + e^{i\xi(t)}\ket{\down}_{A}\ket{\up}_{B}\Bigr]\ket{0}_{A}\ket{0}_{B}.
\end{equation}

%%%%%%%%%%%%%%%%%%%%%%%%%%%%%%%%%%%%%%%%%%%%%%%%%%%%%%%%%%%%%%%%%%%%%%%%%%%%%%%%%%%%%%%%%%%%%%%%%%
%\section*{Mapping}%%%%%%%%%%%%%%%%%%%%%%%%%%%%%%%%%%%%%%%%%%%%%%%%%%%%%%%%%%%%%%%%%%%%%%%%%%%%%%
%%%%%%%%%%%%%%%%%%%%%%%%%%%%%%%%%%%%%%%%%%%%%%%%%%%%%%%%%%%%%%%%%%%%%%%%%%%%%%%%%%%%%%%%%%%%%%%%%

We transfer the entanglement from the spin to the motion with a sequence of laser pulses on the \Beplus ions. Carrier transitions (labeled with superscript \textit{c}, duration $\simeq$ 4 $\mu$s) only affect the spin states, and sideband transitions (superscript \textit{m}, referred to as spin $\leftrightarrow$ motion transfer pulses, duration $\simeq$ 13 $\mu$s) couple the spin and motion. These can be described as generalized rotations:
\[R^{c,m}_{j}(\theta,\phi) = \left( \begin{array}{cc}
 \text{cos}\frac{ \theta}{ 2} & -i\text{e}^{-i\phi}\text{sin}\frac{\theta}{2}  \\
 -i\text{e}^{i\phi}\text{sin}\frac{\theta}{2} & \text{cos}\frac{\theta}{2}
\end{array} \right),
\]
where $j \in \{A, B\}$. Carrier transitions correspond to rotations in the basis
\[\left( \begin{array}{c}
 1\\
 0 \\
\end{array} \right)= \ket{\up} ,
\left( \begin{array}{c}
 0\\
 1 \\
\end{array} \right) = \ket{\down},
\]
and sideband transitions correspond to rotations in the basis
\[ \left( \begin{array}{c}
 1\\
 0 \\
\end{array} \right)= \ket{\up}\ket{1},
\left( \begin{array}{c}
 0\\
 1 \\
\end{array} \right)= \ket{\down}\ket{0} .
\]

\noindent The rotation angle $\theta$ is proportional to the intensity and duration of the pulses, and the phase $\phi$ is determined by the phase difference between the two optical Raman fields\cite{bible, king98} at the position of the ion. We individually address the \Beplus ions in each well using acousto-optic modulators to shift the positions of the laser beams.

Applying $R^{m}_A(\pi,0)$ to state (\ref{state3}) entangles the \Beplus -- \Mgplus motion in well $A$ with the \Beplus spin in well $B$, creating the state
\begin{equation}
\label{spinmo}\textstyle{\frac{1}{\sqrt{2}}} \ket{\up}_{A}\Bigl[\ket{\down}_{B}\ket{0}_{A} - ie^{i\xi(t)}\ket{\up}_{B}\ket{1}_{A}\Bigr]\ket{0}_{B}.
\end{equation}
After this spin $\rightarrow$ motion transfer, the spin in well $B$ is sensitive to decoherence from fluctuating magnetic fields. To minimize this effect, we apply a spin-echo pulse\cite{Vandersypen04}, $R^{c}_B(\pi,0)$, $\textit{T}$  $\simeq$ 40 $\mu$s after the previous pulse. After a second delay $\textit{T}$, we apply a second spin $\rightarrow$ motion transfer pulse $R^{m}_{B}(\pi,0)$ in well $B$, producing the state
\begin{equation}
\label{moent}\textstyle{\frac{1}{\sqrt{2}}} \ket{\up}_{A}\ket{\up}_{B}\Bigl[\ket{0}_{A}\ket{0}_{B}- e^{i\xi(t)}\ket{1}_{A}\ket{1}_{B}\Bigr] .
\end{equation}
This state is an entangled superposition of both stretch modes in the ground and first excited states. The entanglement now resides only in the mechanical oscillator states of both wells. We leave the system in this state for $\sim$ 50 $\mu$s before beginning our analysis.

We are not able to directly measure the entangled motional state. The analysis proceeds by basically reversing the steps used to create  state (\ref{moent}) and characterizing the resulting spin state. We transfer the motional state back into the spins using the pulse sequence: $R^{m}_B(\pi,0), T, R^{c}_B(\pi,0), T, R^{m}_A(\pi,\phi_{A})$. We then recombine all the ions into a single potential well, to ideally reproduce the state $\ket{\Psi^{+}}$, having chosen $\phi_A$ to compensate for the phase $\xi(t)$.

Imperfect creation of the state (\ref{moent}) could leave entanglement in the spin states, which could mimic motional entanglement in the analysis. To prevent this, we transfer residual populations $\epsilon_{A, B}$ of states $\ket{\downarrow}_{A,B}$ into auxiliary internal (hyperfine) states  prior to performing the motion $\rightarrow$ spin transfers (Methods). This residual population does not enhance the deduced entanglement.

%%%%%%%%%%%%%%%%%%%%%%%%%%%%%%%%%%%%%%%%%%%%%%%%%%%%%%%%%%%%%%%%%%%%%%%%%%%%%%%%%%%%%%%%%%%%%%%%%%%%%%%%%%%%%%%%%%%%%%%%%%%%%
%\section*{Analysis}%%%%%%%%%%%%%%%%%%%%%%%%%%%%%%%%%%%%%%%%%%%%%%%%%%%%%%%%%%%%%%%%%%%%%%%%%%%%%%%%%%%%%%%%%%%%%%%%%%%%%%%%%%
%%%%%%%%%%%%%%%%%%%%%%%%%%%%%%%%%%%%%%%%%%%%%%%%%%%%%%%%%%%%%%%%%%%%%%%%%%%%%%%%%%%%%%%%%%%%%%%%%%%%%%%%%%%%%%%%%%%%%%%%%%%%%

\begin{figure}
  \includegraphics[width=\columnwidth]{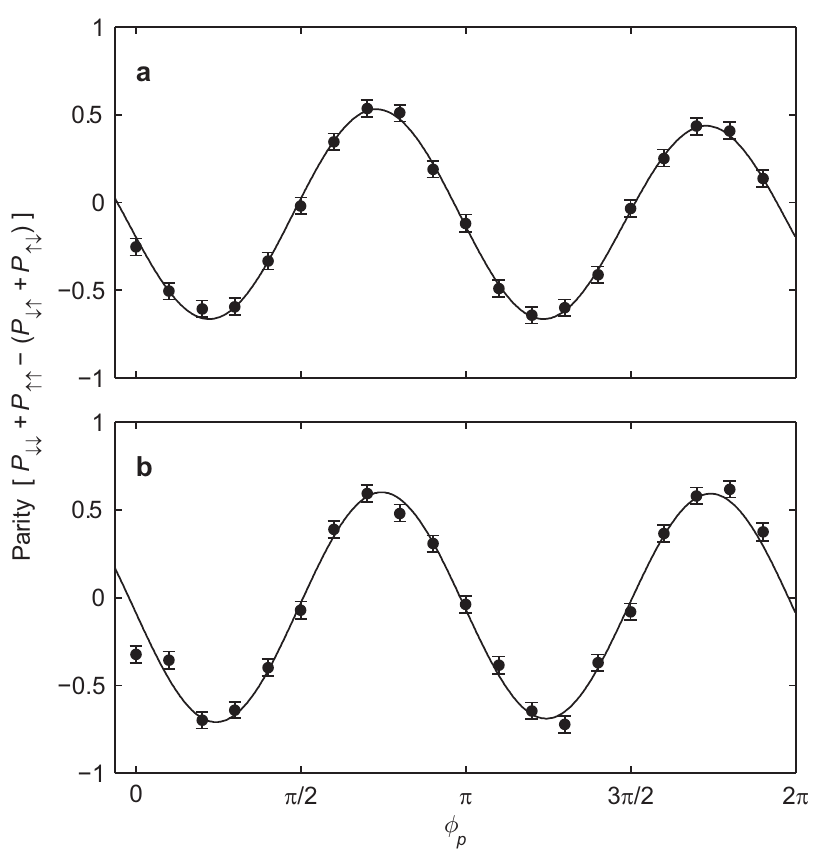}\\
    \caption{\textbf{Entanglement demonstration through parity oscillation.} Parity data obtained from \textbf{a}, the entangled mechanical oscillators and \textbf{b}, the spin -- motion entanglement experiments. Each point is calculated using the maximum-likelihood method on the fluorescence data from running the experiment 500 times. The solid curve is a fit to the data. Two-ion entanglement is verified by an amplitude greater than 0.5 for the component of the parity  signal that oscillates at twice the analysis pulse phase $\phi_{p}$ \cite{sackett00}. For the data shown, this amplitude is \textbf{a}, 0.57(2) and \textbf{b}, 0.65(2).}
\end{figure}

Our detection relies on analyzing the state $\ket{\Psi_{f}} =\frac{1}{\sqrt{2}}[\ket{\upup} + i\ket{\downdown}]$, which we create by applying a common rotation $R^{c}(\frac{\pi}{2},-\frac{3\pi}{4})$ to both spins. We verify the entanglement created in state (\ref{moent}) by measuring the off-diagonal element $|\rho_{\downdown,\upup}| = |\bra{\downdown}\rho_{f}\ket{\upup}|$ of the density matrix $\rho_{f}$ corresponding to our approximation to the state $\ket{\Psi_{f}}$. We determine $|\rho_{\downdown,\upup}|$ by applying a final analysis pulse, $R^{c}(\frac{\pi}{2},\phi_{p})$, to both \Beplus ions with a phase $\phi_{p}$ and measuring the parity\cite{sackett00}, $ P_{\downdown}+P_{\upup} - (P_{\downup}+P_{\updown})$, for different values of $\phi_{p}$, where $P_{\downdown},P_{\upup},P_{\downup},$ and $P_{\updown}$ are the populations of the spin states $\ket{\downdown},\ket{\upup},\ket{\downup},$ and $\ket{\updown}$. The entanglement is revealed by the component of the parity signal that oscillates as  $C_2 \cos(2 \phi_{p})$, where $C_2 = |\rho_{\downdown,\upup}| $. A value of $C_2 > 0.5$ verifies the spin entanglement of $\ket{\Psi_{f}}$ and thus the motional entanglement in state (\ref{moent}).

To deduce the spin populations, we use state-dependent resonance fluorescence\cite{bible,08NatureInsight}. The $\ket{\uparrow}$ state strongly fluoresces. Prior to measurement we transfer the $\ket{\downarrow}$ population to a ``dark" auxiliary state (Methods). The populations $\epsilon_{A, B}$ are in another dark auxiliary state, where they falsely contribute to $P_{\downdown}$ but in a way that does not depend on $\phi_{p}$. We fit the data in Fig. 2a with $C_{2}\text{cos}(2\phi_{p} + \phi_{2}) + C_{1}\text{cos}(\phi_{p} + \phi_1 ) + C_0$ and extract $C_2 = 0.57(2)$. This demonstrates that entanglement was present in the motion after the steps to create state (\ref{moent}).

%%%%%%%%%%%%%%%%%%%%%%%%%%%%%%%%%%%%%%%%%%%%%%%%%%%%%%%%%%%%%%%%%%%%%%%%%%%%%%%%%%%%%%%%%%%%%%%%%%%%%%%%%%%%%%%%%%%%%%%%%
%\section*{Spin-Motion Entanglement}%%%%%%%%%%%%%%%%%%%%%%%%%%%%%%%%%%%%%%%%%%%%%%%%%%%%%%%%%%%%%%%%%%%%%%%%%%%%%%%%%%%%
%%%%%%%%%%%%%%%%%%%%%%%%%%%%%%%%%%%%%%%%%%%%%%%%%%%%%%%%%%%%%%%%%%%%%%%%%%%%%%%%%%%%%%%%%%%%%%%%%%%%%%%%%%%%%%%%%%%%%%%%

The intermediate state (\ref{spinmo}) is itself a novel ``spin -- motion" entangled state, where the spin state of the \Beplus ion in well $B$ is entangled with the motion of the stretch mode of the ion pair in well $A$. We characterize this state in a separate set of experiments. After creating state (\ref{spinmo}), we allow it to persist for 176 $\mu$s. Following the analysis described above (omitting the spin $\leftrightarrow$ motion transfer steps in well B), we measure the parity (Fig. 2b) and find $C_2 = 0.65(2)$.

%%%%%%%%%%%%%%%%%%%%%%%%%%%%%%%%%%%%%%%%%%%%%%%%%%%%%%%%%%%%%%%%%%%%%%%%%%%%%%%%%%%%%%%%%%%%%%%%%%%%%%%%%%%%%%%%%%%%%%%%%%%%
%\section*{Errors}%%%%%%%%%%%%%%%%%%%%%%%%%%%%%%%%%%%%%%%%%%%%%%%%%%%%%%%%%%%%%%%%%%%%%%%%%%%%%%%%%%%%%%%%%%%%%%%%%%%%%%%%%%%
%%%%%%%%%%%%%%%%%%%%%%%%%%%%%%%%%%%%%%%%%%%%%%%%%%%%%%%%%%%%%%%%%%%%%%%%%%%%%%%%%%%%%%%%%%%%%%%%%%%%%%%%%%%%%%%%%%%%%%%%%%%%%

Significant sources of infidelity are spontaneous photon scattering\cite{Ozeri07} and motional decoherence. The fidelity with which we initially create $\ket{\Psi^+}$ is $\simeq $ 0.88. Motional state superpositions of the stretch mode in each well were independently measured to have a coherence time of $\sim$ 800 $\mu$s, which is consistent with a model of decoherence due to couplings to thermally occupied radial modes\cite{Roos08}. In the entangled mechanical oscillators experiment, the motional superpositions are occupied for $\simeq$ 250 $\mu$s and $\simeq$ 50 $\mu$s in wells $A$ and $B$ respectively; we estimate a decrease in $C_2$ from this source to be $\sim$ 5 $\%$. In the spin -- motion entanglement experiment, we estimate a decrease in $C_2$ of $\sim$ 3 $\%$ from this source. Non-zero temperature also reduced the fidelity of motional state initialization. We estimate that this would reduce the fidelity for producing the states (\ref{moent}) and (\ref{spinmo}) by 8 $\%$ and 6 $\%$ respectively. Intensity fluctuations at the few percent level reduce the accuracy of all rotations.

The Coulomb coupling between the ion pairs in wells $A$ and $B$ could lead to an entangled state of their stretch modes. However, the resonant exchange rate is 5 Hz, which leads to negligible entanglement for the experimental time scales.  Furthermore, the stretch mode frequencies in wells $A$ and $B$ differ by $\sim$ 25 kHz, which would yield negligible entanglement for all time scales.

%%%%%%%%%%%%%%%%%%%%%%%%%%%%%%%%%%%%%%%%%%%%%%%%%%%%%%%%%%%%%%%%%%%%%%%%%%%%%%%%%%%%%%%%%%%%%%%%%%%%%%%%%%%%%%%%%%%%%%%%%%%%%
%\section*{Conclusion}%%%%%%%%%%%%%%%%%%%%%%%%%%%%%%%%%%%%%%%%%%%%%%%%%%%%%%%%%%%%%%%%%%%%%%%%%%%%%%%%%%%%%%%%%%%%%%%%%%%%%%%%
%%%%%%%%%%%%%%%%%%%%%%%%%%%%%%%%%%%%%%%%%%%%%%%%%%%%%%%%%%%%%%%%%%%%%%%%%%%%%%%%%%%%%%%%%%%%%%%%%%%%%%%%%%%%%%%%%%%%%%%%

In summary, we have created two novel entangled states of separated systems involving mechanical oscillators, extending the regime where entanglement has been observed in nature. Implementing these experiments required deterministic ion ordering and the ability to separate and recool ions while preserving entanglement and performing subsequent coherent operations. This is the first demonstration of these techniques combined. Some of these methods could apply to similar experiments with nano- and micro-mechanical resonators\cite{05Schwab,08Kippenburg,Mancini02}. The states created could be used to extend tests of nonlocality in ion traps in a manner analogous to that proposed for the electromagnetic fields of separated cavities\cite{Haroche05}. The control developed for these experiments also represents an important step towards large-scale trapped-ion quantum information processing\cite{02Kielpinski,bible}.

\section{Methods}
The methods section details 1) \Mgplus laser cooling of the motional modes of the two multi-species ion configurations, 2) protocols used for transferring population that was not mapped into the entangled motional state and transferring the final $\ket{\downarrow}$ population to a dark state, and 3) two control experiments. The online supplementary table contains more details of the experimental sequence.
\subsection{\Mgplusbf laser cooling of motional modes.}
In addition to comprising part of the mechanical oscillators, the \Mgplus ions serve as a tool to provide sympathetic cooling of the \Beplus ions\cite{barrett03}. Doppler cooling of \Mgplus is accomplished by driving transitions between the ground $^{2}S_{1/2}$ states and excited $^{2}P_{1/2}$ states, which have a radiative linewidth of 41 MHz\cite{Herrmann09}. In the 0.012 T applied magnetic field, the ground Zeeman states $\ket{m_J = \pm 1/2}$ are split by 334 MHz, hence efficient Doppler cooling requires an additional repump beam to prevent optical pumping. One cycle of the pulsed \Mgplus sideband cooling\cite{barrett03} uses stimulated-Raman transitions on a motional sideband of the $\ket{+ 1/2} \rightarrow \ket{-1/2} $ ground state transition (duration $\sim 5$ $\mu$s), followed by application of the repumping beam to reprepare $\ket{+1/2}$ ($\sim 2$ $\mu$s).

At the start of each experiment, the four ions are located in one well in the configuration \Beplus--\Mgplus--\Mgplus--\Beplus, which has four axial modes of motion. In order of ascending frequency, these are the in-phase mode (frequency $\simeq$ 2.0 MHz, mode vector: [0.32, 0.63, 0.63, 0.32]), the out-of-phase mode (4.1 MHz, [-0.47, -0.53, 0.53, 0.47]), a third mode (5.5 MHz, [0.63, -0.32, -0.32, 0.63]) and a fourth mode (5.7 MHz, [0.53, -0.47, 0.47, -0.53]). The amplitudes given in the mode vectors (written in ion order from left to right) are related to each ion's root-mean-squared ground state wavefunction size by multiplying by $\sqrt{\hbar/(2 M \omega)}$, with $M$ the mass of the relevant ion and $\omega$ the mode frequency in angular units. There are also radial modes that have small amplitudes for \Beplus but large amplitudes for \Mgplus. This means \Beplus cooling is inefficient for these modes, hence we also cool these modes using \Mgplus Doppler cooling. After preparing the four axial modes to near the ground state, the geometric phase gate\cite{didi_gate} operation implements a \Beplus state-dependent motional displacement on the out-of-phase mode.

Sympathetic cooling plays a crucial role in making the transition from state (\ref{state2}) to state (\ref{state3}). After separating the ion pairs into wells $A$ and $B$, we simultaneously cool them using \Mgplus Doppler cooling. This is followed by 40 cooling cycles per mode on the second motional sideband and then 60 cycles per mode on the first sideband to prepare the axial modes to near the ground state. The motional modes of each \Beplus -- \Mgplus pair are the ``common'' mode (frequency $\simeq$ 2.3 MHz, mode vector $\simeq$ [0.37, 0.93]) and the ``stretch'' mode (4.9 MHz, [-0.93, 0.37]).

\subsection{Transfer to auxiliary hyperfine states.}
Prior to final spin population measurement, the \ket{\downarrow} population is transferred to the dark state $\ket{F = 2,m_{F}= -2}$ using carrier $\pi$ pulses $R(\pi,0)$ on the sequence of transitions $\ket{2, 1} \rightarrow  \ket{2, 0}$, $\ket{2, 0} \rightarrow  \ket{2, -1}$, $\ket{2, -1} \rightarrow \ket{2, -2}$. The number of photons  measured per \Beplus ion if all the population were in the $\ket{2, -2}$ dark state during the 200 $\mu$s detection period approximates a Poisson distribution with a mean of 0.2. For the fluorescing state $\ket{\uparrow}$, we observe a Poisson distribution, with a mean number of photons $\simeq$ 10 per \Beplus ion.

As described in the main text, we move populations $\epsilon_{A, B}$ of the spin states $\ket{\downarrow}_{A,B}$ to the auxiliary hyperfine state $\ket{2, 0}$, so they do not contribute to entanglement verification. To ensure that the $\epsilon_{A, B}$ populations end in dark states for the measurements, we precede the transfer pulses described in the previous paragraph with transfer of the $\ket{2,0}$ populations to $\ket{2, -2}$ using a sequence of carrier $\pi$ pulses on the $\ket{2, 0} \rightarrow  \ket{2, -1}$ and  $\ket{2, -1} \rightarrow \ket{2, -2}$ transitions. Since the last pulse of the final transfer sequence is also a carrier $\pi$ pulse on the $\ket{2, -2} \leftrightarrow \ket{2,-1}$ transition, this leads to the populations $\epsilon_{A, B}$ ending in $\ket{2, -1}$. If all the population is in this state, it would give a mean fluorescence value per \Beplus ion of $\sim 1$ photon during detection. This fluorescence is independent of the final analysis pulse phase $\phi_p$, and hence does not contribute to $C_2$.

\subsection{Control experiments.}To provide partial checks of the spin $\rightarrow$ motion transfer steps, we perform separate experiments to determine the spin populations after the transfer. In the first check experiment, we follow the steps used to create state (\ref{spinmo}) then implement the above hyperfine state transfer sequences (omitting the $\epsilon_{B}$ population transfer process) and measure the spin populations. The populations are determined to be $P_{\uparrow \uparrow} = 0.47(1), P_{\downarrow \downarrow} = 0.04(1)  ,$ and $P_{\downarrow \uparrow} + P_{\uparrow \downarrow} = 0.49(2) $. Ideally we would expect  $P_{\uparrow \uparrow} = 1/2, P_{\downarrow \downarrow} = 0  ,$ and $P_{\downarrow \uparrow} + P_{\uparrow \downarrow} =1/2 $. Similarly, after the step used to create state (\ref{moent}), and following the transfer procedure, we determine $P_{\uparrow \uparrow} = 0.86(2), P_{\downarrow \downarrow} = 0.01(1)  ,$ and $P_{\downarrow \uparrow} + P_{\uparrow \downarrow} = 0.13(2)  $. Ideally we should find $P_{\uparrow \uparrow} = 1$.

\begin{acknowledgments}
This work was supported by IARPA and the NIST Quantum Information Program. J. P. Home acknowledges support from the Lindemann Trust. We thank J. Britton, Y. Colombe, and H. Uys for comments on the manuscript. This paper is a contribution by the National Institute of Standards and Technology and not subject to U.S. copyright.  The authors declare that they have no competing financial interests.
\end{acknowledgments}

\bibliography{EMO_paper_references_archiv}

\begin{thebibliography}{30}
\expandafter\ifx\csname natexlab\endcsname\relax\def\natexlab#1{#1}\fi
\expandafter\ifx\csname bibnamefont\endcsname\relax
  \def\bibnamefont#1{#1}\fi
\expandafter\ifx\csname bibfnamefont\endcsname\relax
  \def\bibfnamefont#1{#1}\fi
\expandafter\ifx\csname citenamefont\endcsname\relax
  \def\citenamefont#1{#1}\fi
\expandafter\ifx\csname url\endcsname\relax
  \def\url#1{\texttt{#1}}\fi
\expandafter\ifx\csname urlprefix\endcsname\relax\def\urlprefix{URL }\fi
\providecommand{\bibinfo}[2]{#2}
\providecommand{\eprint}[2][]{\url{#2}}

\bibitem[{\citenamefont{Schr\"{o}dinger}(1935)}]{Schrodinger1935}
\bibinfo{author}{\bibfnamefont{E.}~\bibnamefont{Schr\"{o}dinger}},
  \bibinfo{journal}{Naturwissenschaften} \textbf{\bibinfo{volume}{23}},
  \bibinfo{pages}{807} (\bibinfo{year}{1935}).

\bibitem[{\citenamefont{Ball}(2008)}]{08Ball}
\bibinfo{author}{\bibfnamefont{P.}~\bibnamefont{Ball}},
  \bibinfo{journal}{Nature} \textbf{\bibinfo{volume}{453}}, \bibinfo{pages}{22}
  (\bibinfo{year}{2008}).

\bibitem[{\citenamefont{Schlosshauer}(2008)}]{08Schlosshauer}
\bibinfo{author}{\bibfnamefont{M.}~\bibnamefont{Schlosshauer}},
  \bibinfo{journal}{Nature} \textbf{\bibinfo{volume}{453}}, \bibinfo{pages}{39}
  (\bibinfo{year}{2008}).

\bibitem[{\citenamefont{Bassi and Ghiradi}(2003)}]{03Bassi}
\bibinfo{author}{\bibfnamefont{A.}~\bibnamefont{Bassi}} \bibnamefont{and}
  \bibinfo{author}{\bibfnamefont{G.}~\bibnamefont{Ghiradi}},
  \bibinfo{journal}{Phys. Rep.} \textbf{\bibinfo{volume}{379}},
  \bibinfo{pages}{257} (\bibinfo{year}{2003}).

\bibitem[{\citenamefont{Leggett}(2002)}]{02Leggett}
\bibinfo{author}{\bibfnamefont{A.~J.} \bibnamefont{Leggett}},
  \bibinfo{journal}{J. Phys.: Condens. Matter.} \textbf{\bibinfo{volume}{14}},
  \bibinfo{pages}{R415} (\bibinfo{year}{2002}).

\bibitem[{\citenamefont{Aspelmeyer and Zeilinger}(2008)}]{Zeilinger08}
\bibinfo{author}{\bibfnamefont{M.}~\bibnamefont{Aspelmeyer}} \bibnamefont{and}
  \bibinfo{author}{\bibfnamefont{A.}~\bibnamefont{Zeilinger}},
  \bibinfo{journal}{Physics World} \textbf{\bibinfo{volume}{7}},
  \bibinfo{pages}{22} (\bibinfo{year}{2008}).

\bibitem[{08N(2008)}]{08NatureInsight}
\bibinfo{journal}{Nature} \textbf{\bibinfo{volume}{453}}, \bibinfo{pages}{1003}
  (\bibinfo{year}{2008}).

\bibitem[{\citenamefont{Mancini et~al.}(2002)\citenamefont{Mancini,
  Giovannetti, Vitali, and Tombesi}}]{Mancini02}
\bibinfo{author}{\bibfnamefont{S.}~\bibnamefont{Mancini}},
  \bibinfo{author}{\bibfnamefont{V.}~\bibnamefont{Giovannetti}},
  \bibinfo{author}{\bibfnamefont{D.}~\bibnamefont{Vitali}}, \bibnamefont{and}
  \bibinfo{author}{\bibfnamefont{P.}~\bibnamefont{Tombesi}},
  \bibinfo{journal}{Phys. Rev. Lett.} \textbf{\bibinfo{volume}{88}},
  \bibinfo{pages}{120401} (\bibinfo{year}{2002}).

\bibitem[{\citenamefont{Schwab and Rourkes}(2005)}]{05Schwab}
\bibinfo{author}{\bibfnamefont{K.~C.} \bibnamefont{Schwab}} \bibnamefont{and}
  \bibinfo{author}{\bibfnamefont{M.~L.} \bibnamefont{Rourkes}},
  \bibinfo{journal}{Physics Today} \textbf{\bibinfo{volume}{58}},
  \bibinfo{pages}{7, 36} (\bibinfo{year}{2005}).

\bibitem[{\citenamefont{Kippenberg and Vahala}(2008)}]{08Kippenburg}
\bibinfo{author}{\bibfnamefont{T.~J.} \bibnamefont{Kippenberg}}
  \bibnamefont{and} \bibinfo{author}{\bibfnamefont{K.~J.}
  \bibnamefont{Vahala}}, \bibinfo{journal}{Science}
  \textbf{\bibinfo{volume}{321}}, \bibinfo{pages}{1172} (\bibinfo{year}{2008}).

\bibitem[{\citenamefont{Milman et~al.}(2005)\citenamefont{Milman, Auffeves,
  Yamaguchi, Brune, Raimond, and Haroche}}]{Haroche05}
\bibinfo{author}{\bibfnamefont{P.}~\bibnamefont{Milman}},
  \bibinfo{author}{\bibfnamefont{A.}~\bibnamefont{Auffeves}},
  \bibinfo{author}{\bibfnamefont{F.}~\bibnamefont{Yamaguchi}},
  \bibinfo{author}{\bibfnamefont{M.}~\bibnamefont{Brune}},
  \bibinfo{author}{\bibfnamefont{J.}~\bibnamefont{Raimond}}, \bibnamefont{and}
  \bibinfo{author}{\bibfnamefont{S.}~\bibnamefont{Haroche}},
  \bibinfo{journal}{Eur. Phys. J. D} \textbf{\bibinfo{volume}{32}},
  \bibinfo{pages}{233} (\bibinfo{year}{2005}).

\bibitem[{\citenamefont{Wineland et~al.}(1998)\citenamefont{Wineland, Monroe,
  Itano, Leibfried, King, and Meekhof}}]{bible}
\bibinfo{author}{\bibfnamefont{D.~J.} \bibnamefont{Wineland}},
  \bibinfo{author}{\bibfnamefont{C.}~\bibnamefont{Monroe}},
  \bibinfo{author}{\bibfnamefont{W.~M.} \bibnamefont{Itano}},
  \bibinfo{author}{\bibfnamefont{D.}~\bibnamefont{Leibfried}},
  \bibinfo{author}{\bibfnamefont{B.~E.} \bibnamefont{King}}, \bibnamefont{and}
  \bibinfo{author}{\bibfnamefont{D.~M.} \bibnamefont{Meekhof}},
  \bibinfo{journal}{J. Res. Natl. Inst. Stand. Tech.}
  \textbf{\bibinfo{volume}{103}}, \bibinfo{pages}{259} (\bibinfo{year}{1998}).

\bibitem[{\citenamefont{Cirac and Zoller}(2000)}]{Cirac00}
\bibinfo{author}{\bibfnamefont{J.~I.} \bibnamefont{Cirac}} \bibnamefont{and}
  \bibinfo{author}{\bibfnamefont{P.}~\bibnamefont{Zoller}},
  \bibinfo{journal}{Nature} \textbf{\bibinfo{volume}{404}},
  \bibinfo{pages}{579} (\bibinfo{year}{2000}).

\bibitem[{\citenamefont{Kielpinski et~al.}(2002)\citenamefont{Kielpinski,
  Monroe, and Wineland}}]{02Kielpinski}
\bibinfo{author}{\bibfnamefont{D.}~\bibnamefont{Kielpinski}},
  \bibinfo{author}{\bibfnamefont{C.}~\bibnamefont{Monroe}}, \bibnamefont{and}
  \bibinfo{author}{\bibfnamefont{D.}~\bibnamefont{Wineland}},
  \bibinfo{journal}{Nature} \textbf{\bibinfo{volume}{417}},
  \bibinfo{pages}{709} (\bibinfo{year}{2002}).

\bibitem[{\citenamefont{Schleich}(2001)}]{Schleich01}
\bibinfo{author}{\bibfnamefont{W.~P.} \bibnamefont{Schleich}},
  \emph{\bibinfo{title}{Quantum Optics in Phase Space}}
  (\bibinfo{publisher}{Wiley-VCH}, \bibinfo{address}{Berlin},
  \bibinfo{year}{2001}), \bibinfo{edition}{1st} ed.

\bibitem[{\citenamefont{Aspect}(2002)}]{Aspect02}
\bibinfo{author}{\bibfnamefont{A.}~\bibnamefont{Aspect}}, in
  \emph{\bibinfo{booktitle}{Quantum [Un]speakables - From Bell to Quantum
  Information}}, edited by \bibinfo{editor}{\bibfnamefont{R.~A.}
  \bibnamefont{Bertlmann}} \bibnamefont{and}
  \bibinfo{editor}{\bibfnamefont{A.}~\bibnamefont{Zeilinger}}
  (\bibinfo{publisher}{Springer-Verlag Berlin Heidelberg},
  \bibinfo{address}{Germany}, \bibinfo{year}{2002}).

\bibitem[{\citenamefont{Matsukevich et~al.}(2008)\citenamefont{Matsukevich,
  Maunz, Moehring, Olmschenk, and Monroe}}]{Matsukevich08}
\bibinfo{author}{\bibfnamefont{D.~N.} \bibnamefont{Matsukevich}},
  \bibinfo{author}{\bibfnamefont{P.}~\bibnamefont{Maunz}},
  \bibinfo{author}{\bibfnamefont{D.~L.} \bibnamefont{Moehring}},
  \bibinfo{author}{\bibfnamefont{S.}~\bibnamefont{Olmschenk}},
  \bibnamefont{and} \bibinfo{author}{\bibfnamefont{C.}~\bibnamefont{Monroe}},
  \bibinfo{journal}{Phys. Rev. Lett.} \textbf{\bibinfo{volume}{100}},
  \bibinfo{pages}{150404} (\bibinfo{year}{2008}).

\bibitem[{\citenamefont{Pan et~al.}(2008)\citenamefont{Pan, Chen, Zukowski,
  Weinfurter, and Zeilinger}}]{08Pan}
\bibinfo{author}{\bibfnamefont{J.-W.} \bibnamefont{Pan}},
  \bibinfo{author}{\bibfnamefont{Z.-B.} \bibnamefont{Chen}},
  \bibinfo{author}{\bibfnamefont{M.}~\bibnamefont{Zukowski}},
  \bibinfo{author}{\bibfnamefont{H.}~\bibnamefont{Weinfurter}},
  \bibnamefont{and}
  \bibinfo{author}{\bibfnamefont{A.}~\bibnamefont{Zeilinger}},
  \bibinfo{journal}{arxiv:0805.2853}  (\bibinfo{year}{2008}).

\bibitem[{\citenamefont{Leibfried et~al.}(2003)\citenamefont{Leibfried,
  DeMarco, Meyer, Lucas, Barrett, Britton, Itano, Jelenkovi{\'c}, Langer,
  Rosenband et~al.}}]{didi_gate}
\bibinfo{author}{\bibfnamefont{D.}~\bibnamefont{Leibfried}},
  \bibinfo{author}{\bibfnamefont{B.}~\bibnamefont{DeMarco}},
  \bibinfo{author}{\bibfnamefont{V.}~\bibnamefont{Meyer}},
  \bibinfo{author}{\bibfnamefont{D.}~\bibnamefont{Lucas}},
  \bibinfo{author}{\bibfnamefont{M.}~\bibnamefont{Barrett}},
  \bibinfo{author}{\bibfnamefont{J.}~\bibnamefont{Britton}},
  \bibinfo{author}{\bibfnamefont{W.~M.} \bibnamefont{Itano}},
  \bibinfo{author}{\bibfnamefont{B.}~\bibnamefont{Jelenkovi{\'c}}},
  \bibinfo{author}{\bibfnamefont{C.}~\bibnamefont{Langer}},
  \bibinfo{author}{\bibfnamefont{T.}~\bibnamefont{Rosenband}},
  \bibnamefont{et~al.}, \bibinfo{journal}{Nature}
  \textbf{\bibinfo{volume}{422}}, \bibinfo{pages}{412} (\bibinfo{year}{2003}).

\bibitem[{\citenamefont{Rowe et~al.}(2002)\citenamefont{Rowe, Ben-Kish,
  De{M}arco, Leibfried, Meyer, Beall, Britton, Hughes, Itano, Jelenkovi{\'c}
  et~al.}}]{rowe02}
\bibinfo{author}{\bibfnamefont{M.~A.} \bibnamefont{Rowe}},
  \bibinfo{author}{\bibfnamefont{A.}~\bibnamefont{Ben-Kish}},
  \bibinfo{author}{\bibfnamefont{B.}~\bibnamefont{De{M}arco}},
  \bibinfo{author}{\bibfnamefont{D.}~\bibnamefont{Leibfried}},
  \bibinfo{author}{\bibfnamefont{V.}~\bibnamefont{Meyer}},
  \bibinfo{author}{\bibfnamefont{J.}~\bibnamefont{Beall}},
  \bibinfo{author}{\bibfnamefont{J.}~\bibnamefont{Britton}},
  \bibinfo{author}{\bibfnamefont{J.}~\bibnamefont{Hughes}},
  \bibinfo{author}{\bibfnamefont{W.~M.} \bibnamefont{Itano}},
  \bibinfo{author}{\bibfnamefont{B.}~\bibnamefont{Jelenkovi{\'c}}},
  \bibnamefont{et~al.}, \bibinfo{journal}{Quant. Inf. Comp.}
  \textbf{\bibinfo{volume}{2}}, \bibinfo{pages}{257} (\bibinfo{year}{2002}).

\bibitem[{\citenamefont{Barrett et~al.}(2004)\citenamefont{Barrett, Chiaverini,
  Sch{\"a}tz, Britton, Itano, Jost, Knill, Langer, Leibfried, Ozeri
  et~al.}}]{barrett04}
\bibinfo{author}{\bibfnamefont{M.~D.} \bibnamefont{Barrett}},
  \bibinfo{author}{\bibfnamefont{J.}~\bibnamefont{Chiaverini}},
  \bibinfo{author}{\bibfnamefont{T.}~\bibnamefont{Sch{\"a}tz}},
  \bibinfo{author}{\bibfnamefont{J.}~\bibnamefont{Britton}},
  \bibinfo{author}{\bibfnamefont{W.~M.} \bibnamefont{Itano}},
  \bibinfo{author}{\bibfnamefont{J.~D.} \bibnamefont{Jost}},
  \bibinfo{author}{\bibfnamefont{E.}~\bibnamefont{Knill}},
  \bibinfo{author}{\bibfnamefont{C.}~\bibnamefont{Langer}},
  \bibinfo{author}{\bibfnamefont{D.}~\bibnamefont{Leibfried}},
  \bibinfo{author}{\bibfnamefont{R.}~\bibnamefont{Ozeri}},
  \bibnamefont{et~al.}, \bibinfo{journal}{Nature}
  \textbf{\bibinfo{volume}{429}}, \bibinfo{pages}{737} (\bibinfo{year}{2004}).

\bibitem[{\citenamefont{Rosenband}()}]{privatetill}
\bibinfo{author}{\bibfnamefont{T.}~\bibnamefont{Rosenband}},
  \bibinfo{note}{private communication}.

\bibitem[{\citenamefont{King et~al.}(1998)\citenamefont{King, Wood, Myatt,
  Turchette, Leibfried, Itano, Monroe, and Wineland}}]{king98}
\bibinfo{author}{\bibfnamefont{B.~E.} \bibnamefont{King}},
  \bibinfo{author}{\bibfnamefont{C.~S.} \bibnamefont{Wood}},
  \bibinfo{author}{\bibfnamefont{C.~J.} \bibnamefont{Myatt}},
  \bibinfo{author}{\bibfnamefont{Q.~A.} \bibnamefont{Turchette}},
  \bibinfo{author}{\bibfnamefont{D.}~\bibnamefont{Leibfried}},
  \bibinfo{author}{\bibfnamefont{W.~M.} \bibnamefont{Itano}},
  \bibinfo{author}{\bibfnamefont{C.}~\bibnamefont{Monroe}}, \bibnamefont{and}
  \bibinfo{author}{\bibfnamefont{D.~J.} \bibnamefont{Wineland}},
  \bibinfo{journal}{Phys. Rev. Lett.} \textbf{\bibinfo{volume}{81}},
  \bibinfo{pages}{1525} (\bibinfo{year}{1998}).

\bibitem[{\citenamefont{Kielpinski et~al.}(2001)\citenamefont{Kielpinski,
  Meyer, Rowe, Sackett, Itano, Monroe, and Wineland}}]{kielpinski01}
\bibinfo{author}{\bibfnamefont{D.}~\bibnamefont{Kielpinski}},
  \bibinfo{author}{\bibfnamefont{V.}~\bibnamefont{Meyer}},
  \bibinfo{author}{\bibfnamefont{M.~A.} \bibnamefont{Rowe}},
  \bibinfo{author}{\bibfnamefont{C.~A.} \bibnamefont{Sackett}},
  \bibinfo{author}{\bibfnamefont{W.~M.} \bibnamefont{Itano}},
  \bibinfo{author}{\bibfnamefont{C.}~\bibnamefont{Monroe}}, \bibnamefont{and}
  \bibinfo{author}{\bibfnamefont{D.~J.} \bibnamefont{Wineland}},
  \bibinfo{journal}{Science} \textbf{\bibinfo{volume}{291}},
  \bibinfo{pages}{1013} (\bibinfo{year}{2001}).

\bibitem[{\citenamefont{Barrett et~al.}(2003)\citenamefont{Barrett, De{M}arco,
  Schaetz, Meyer, Leibfried, Britton, Chiaverini, Itano, Jelenkovi{\'c}, Jost
  et~al.}}]{barrett03}
\bibinfo{author}{\bibfnamefont{M.~D.} \bibnamefont{Barrett}},
  \bibinfo{author}{\bibfnamefont{B.}~\bibnamefont{De{M}arco}},
  \bibinfo{author}{\bibfnamefont{T.}~\bibnamefont{Schaetz}},
  \bibinfo{author}{\bibfnamefont{V.}~\bibnamefont{Meyer}},
  \bibinfo{author}{\bibfnamefont{D.}~\bibnamefont{Leibfried}},
  \bibinfo{author}{\bibfnamefont{J.}~\bibnamefont{Britton}},
  \bibinfo{author}{\bibfnamefont{J.}~\bibnamefont{Chiaverini}},
  \bibinfo{author}{\bibfnamefont{W.~M.} \bibnamefont{Itano}},
  \bibinfo{author}{\bibfnamefont{B.}~\bibnamefont{Jelenkovi{\'c}}},
  \bibinfo{author}{\bibfnamefont{J.~D.} \bibnamefont{Jost}},
  \bibnamefont{et~al.}, \bibinfo{journal}{Phys. Rev. A}
  \textbf{\bibinfo{volume}{68}}, \bibinfo{pages}{042302}
  (\bibinfo{year}{2003}).

\bibitem[{\citenamefont{Vandersypen and Chuang}(2004)}]{Vandersypen04}
\bibinfo{author}{\bibfnamefont{L.~M.~K.} \bibnamefont{Vandersypen}}
  \bibnamefont{and} \bibinfo{author}{\bibfnamefont{I.~L.}
  \bibnamefont{Chuang}}, \bibinfo{journal}{Rev. Mod. Phys.}
  \textbf{\bibinfo{volume}{76}}, \bibinfo{pages}{1037} (\bibinfo{year}{2004}).

\bibitem[{\citenamefont{Sackett et~al.}(2000)\citenamefont{Sackett, Kielpinski,
  King, Langer, Meyer, Myatt, Rowe, Turchette, Itano, Wineland
  et~al.}}]{sackett00}
\bibinfo{author}{\bibfnamefont{C.~A.} \bibnamefont{Sackett}},
  \bibinfo{author}{\bibfnamefont{D.}~\bibnamefont{Kielpinski}},
  \bibinfo{author}{\bibfnamefont{B.~E.} \bibnamefont{King}},
  \bibinfo{author}{\bibfnamefont{C.}~\bibnamefont{Langer}},
  \bibinfo{author}{\bibfnamefont{V.}~\bibnamefont{Meyer}},
  \bibinfo{author}{\bibfnamefont{C.~J.} \bibnamefont{Myatt}},
  \bibinfo{author}{\bibfnamefont{M.}~\bibnamefont{Rowe}},
  \bibinfo{author}{\bibfnamefont{Q.~A.} \bibnamefont{Turchette}},
  \bibinfo{author}{\bibfnamefont{W.~M.} \bibnamefont{Itano}},
  \bibinfo{author}{\bibfnamefont{D.~J.} \bibnamefont{Wineland}},
  \bibnamefont{et~al.}, \bibinfo{journal}{Nature}
  \textbf{\bibinfo{volume}{404}}, \bibinfo{pages}{256} (\bibinfo{year}{2000}).

\bibitem[{\citenamefont{Ozeri et~al.}(2007)\citenamefont{Ozeri, Itano,
  Blakestad, Britton, Chiaverini, Jost, Langer, Leibfried, Reichle, Seidelin
  et~al.}}]{Ozeri07}
\bibinfo{author}{\bibfnamefont{R.}~\bibnamefont{Ozeri}},
  \bibinfo{author}{\bibfnamefont{W.~M.} \bibnamefont{Itano}},
  \bibinfo{author}{\bibfnamefont{R.}~\bibnamefont{Blakestad}},
  \bibinfo{author}{\bibfnamefont{J.}~\bibnamefont{Britton}},
  \bibinfo{author}{\bibfnamefont{J.}~\bibnamefont{Chiaverini}},
  \bibinfo{author}{\bibfnamefont{J.}~\bibnamefont{Jost}},
  \bibinfo{author}{\bibfnamefont{C.}~\bibnamefont{Langer}},
  \bibinfo{author}{\bibfnamefont{D.}~\bibnamefont{Leibfried}},
  \bibinfo{author}{\bibfnamefont{R.}~\bibnamefont{Reichle}},
  \bibinfo{author}{\bibfnamefont{S.}~\bibnamefont{Seidelin}},
  \bibnamefont{et~al.}, \bibinfo{journal}{Phys. Rev. A}
  \textbf{\bibinfo{volume}{75}}, \bibinfo{pages}{042329}
  (\bibinfo{year}{2007}).

\bibitem[{\citenamefont{Roos et~al.}(2008)\citenamefont{Roos, Monz, Kim, Riebe,
  Häffner, James, and Blatt}}]{Roos08}
\bibinfo{author}{\bibfnamefont{C.~F.} \bibnamefont{Roos}},
  \bibinfo{author}{\bibfnamefont{T.}~\bibnamefont{Monz}},
  \bibinfo{author}{\bibfnamefont{K.}~\bibnamefont{Kim}},
  \bibinfo{author}{\bibfnamefont{M.}~\bibnamefont{Riebe}},
  \bibinfo{author}{\bibfnamefont{H.}~\bibnamefont{Häffner}},
  \bibinfo{author}{\bibfnamefont{D.~F.~V.} \bibnamefont{James}},
  \bibnamefont{and} \bibinfo{author}{\bibfnamefont{R.}~\bibnamefont{Blatt}},
  \bibinfo{journal}{Phys. Rev. A} \textbf{\bibinfo{volume}{77}},
  \bibinfo{pages}{040302(R)} (\bibinfo{year}{2008}).

\bibitem[{\citenamefont{Herrmann et~al.}(2009)\citenamefont{Herrmann,
  Batteiger, Knünz, Saathoff, Udem, and Hänsch}}]{Herrmann09}
\bibinfo{author}{\bibfnamefont{M.}~\bibnamefont{Herrmann}},
  \bibinfo{author}{\bibfnamefont{V.}~\bibnamefont{Batteiger}},
  \bibinfo{author}{\bibfnamefont{S.}~\bibnamefont{Knünz}},
  \bibinfo{author}{\bibfnamefont{G.}~\bibnamefont{Saathoff}},
  \bibinfo{author}{\bibfnamefont{T.}~\bibnamefont{Udem}}, \bibnamefont{and}
  \bibinfo{author}{\bibfnamefont{T.~W.} \bibnamefont{Hänsch}},
  \bibinfo{journal}{Phys. Rev. Lett.} \textbf{\bibinfo{volume}{102}},
  \bibinfo{pages}{013006} (\bibinfo{year}{2009}).

\end{thebibliography}

\appendix*
\begin{scriptsize}	% set the font size

%\sffamily		% To make it all sans-serif, as Nature tables typically are (uncomment \usepackage{sfmath} above)

% Set up the caption
\renewcommand{\tablename}{\textbf{Supplementary Table}}	% "Supplementary Table" instead of "Table"
\renewcommand{\thetable}{\textbf{\arabic{table}}}				% Bold table number
\setlength{\LTcapwidth}{5.75in}													% make it wider
% The \LT@makecaption command has been changed in the nature.cls file to do a bold vertical bar instead of a colon.

\newcommand{\ssp}{\renewcommand{\baselinestretch}{1}\scriptsize}	% define a single-spaced, properly sized font column style

\begin{longtable*}{>{\raggedleft\ssp}m{0.9in}|>{\ssp\centering}m{2.5in}|>{\ssp\centering}m{0.7in}|>{\ssp}m{1.75in}}

		\caption[]{\textbf{Detailed procedure for the entangled mechanical oscillators experiment.} Rotations are defined in the text; those without superscripts are carrier rotations between hyperfine state pairs other than $\{\ket{\up},\ket{\down}\}$. The auxiliary states are indicated with $\ket{F,m_F}$. Durations are rounded to the nearest microsecond. Steps without an explicit operation (noted as -- ) typically involve laser frequency changes and intensity stabilization. The procedure for the spin--motion entanglement experiment is identical, except that the seven steps between the spin-echo pulses in well $B$ are replaced with a single 90~$\mu$s delay.} \\
 		\hline\hline
Operation &  Ideal State after Operation 	& Duration ($\mu$s)  & Notes\\
		\hline
\endfirsthead

		\caption[]{(continued)} \\
 		\hline\hline
Operation &  Ideal State after Operation	& Duration ($\mu$s)  & Notes\\
		\hline
\endhead

		\hline \multicolumn{4}{r}{\emph{Continued on next page}}
\endfoot

\endlastfoot

Order ions								& --						& 935		& Order the ions to\newline \Beplus--\Mgplus--\Mgplus--\Beplus 	\\
%Lock the laser intensity	& --						& 380		&	-- \\ % Was lock-up 380 + -- 26. Now -- 406
		--										&	--						&	406		& -- \\
Doppler cool\linebreak(\Beplus \& \Mgplus)
													&	--						& 3500	& -- \\
Doppler cool\linebreak(\Beplus only)
													& --						& 500		& -- \\
		--										& --						& 2			& -- \\
Repump \Mgplus						&	--						& 2			& -- \\
Repump \Beplus						& $\uu$					&	25		& -- \\
\Beplus sideband cool			& $\uu$					& 2753	& Cool the four axial modes to the ground state (20 cooling cycles per mode). \\
Prepare $\ket{\Psi^{+}}$	& $\frac{1}{\sqrt{2}}\Bigl[\ud+\du\Bigr]$
																					& 266		& Use a geometric phase gate and carrier transitions (state(1)) \\
Move and separate					& $\frac{1}{\sqrt{2}}\Bigl[\ket{\up}_{A}\ket{\down}_{B}+e^{i\xi(t)}\ket{\down}_{A}\ket{\up}_{B}\Bigr]$
																					& 819 	& Separate the ions into wells $A$ and $B$. Each well holds a
																											\Beplus--\Mgplus pair in an unknown motional state (state (2)). \\
\Mgplus Doppler cool			&	--						& 400 	& Cool in both wells simultaneously; \Beplus coherence undisturbed \\
\Mgplus second-sideband~cool
													& --						& 1078 	& 40 cooling cycles on the second sideband of each of the two axial modes;
																											cool both wells simultaneously \\
\Mgplus first-sideband~cool
													& $\frac{1}{\sqrt{2}}
															 \Bigl[\ket{\up}_{A}\ket{\down}_{B}+e^{i\xi(t)}\ket{\down}_{A}\ket{\up}_{B}\Bigr]\ket{0}_{A}\ket{0}_{B}$
																					& 1277 	& Account for the slight difference in mode frequencies between
																											the two wells by applying 30 cooling cycles resonant with each mode (4 modes total).
																											Final $\langle n\rangle < 0.1$ on the stretch modes.  (state (3)) \\
		--										& -- 						& 22		& -- \\
$R^{m}_{A}(\pi,0)$ 				& $\frac{1}{\sqrt{2}}\ket{\up}_{A}\Bigl[\ket{\down}_{B}\ket{0}_{A}
																	- ie^{i\xi(t)}\ket{\up}_{B}\ket{1}_{A}\Bigr]\ket{0}_{B}$
																					& 12		& Spin$\rightarrow$motion transfer pulse in well $A$ (state (4)) \\
		--										&	--						&	14 		& \textbf{At this point, we have entangled a mechanical oscillator
																														with the spin of a separated ion.} \\
$R_{A}(\pi,0)$ 						& $\ket{\down}_{A} \rightarrow \ket{2,0}_{A}$
																					& 3			& Transfer residual population $\epsilon_{A}$ of $\ket{\down}_{A}$
																											($\epsilon_{A} \neq 0$ if there is an error in the
																											spin$\rightarrow$motion transfer) \\
		--										& --						& 22		& This step and the previous two constitute $T$ of the first spin-echo sequence. \\
$R^{c}_{B}(\pi,0)$  			& $\frac{1}{\sqrt{2}}\ket{\up}_{A}\Bigl[\ket{\up}_{B}\ket{0}_{A}
																	- ie^{i\xi(t)}\ket{\down}_{B}\ket{1}_{A}\Bigr]\ket{0}_{B}$
																					& 4			& First spin-echo pulse in well $B$  \\
		--										& --						& 38		& Second $T$ delay of the first spin-echo sequence \\
$R^{m}_{B}(\pi,0)$ 				& $\frac{1}{\sqrt{2}}\ket{\up}_{A}\ket{\up}_{B}\Bigl[\ket{0}_{A}\ket{0}_{B}
																	- e^{i\xi(t)}\ket{1}_{A}\ket{1}_{B}\Bigr]$
																					& 14		& Spin$\rightarrow$motion transfer pulse in well $B$ (state (5)) \\
		--										&	--						& 24 		& \textbf{At this point, we have entangled separated mechanical oscillators}. \\
$R_{B}(\pi,0)$ 						& $\ket{\down}_{B} \rightarrow \ket{2,0}_{B}$
																					& 4 		& Transfer residual population $\epsilon_{B}$ of $\ket{\down}_{B}$.
																											As above, this removes any residual spin entanglement
																											for the remainder of the experiment.\\
		--										& --						& 24		& -- \\
$R^{m}_{B}(\pi,0)$ 				& $\frac{1}{\sqrt{2}}\ket{\up}_{A}\Bigl[\ket{\up}_{B}\ket{0}_{A}
																	+ i e^{i\xi(t)}\ket{\down}_{B}\ket{1}_{A}\Bigr]\ket{0}_{B}$
																					& 14	 	& Motion$\rightarrow$spin transfer pulse in well $B$  \\
		--										& --						& 38		& First $T$ delay of the second spin-echo sequence\\
$R^{c}_{B}(\pi,0)$  			& $\frac{1}{\sqrt{2}}\ket{\up}_{A}\Bigl[\ket{\down}_{B}\ket{0}_{A}	
																	+ ie^{i\xi(t)}\ket{\up}_{B}\ket{1}_{A}\Bigr]\ket{0}_{B}$
																					& 4			& Second spin-echo pulse in well $B$ \\
		--										& --						& 39		& Second $T$ delay of the second spin-echo sequence\\
$R^{m}_{A}(\pi,\phi_{A})$ & $\frac{1}{\sqrt{2}}\Bigl[\ket{\up}_{A}\ket{\down}_{B}
																	+ e^{i\left(\xi(t)+ \phi_{A}\right)}\ket{\down}_{A}\ket{\up}_{B}\Bigr]\ket{0}_{A}\ket{0}_{B}$
																					& 11 		& Motion $\rightarrow$ spin transfer pulse in well $A$\\
Recombine 								& $\frac{1}{\sqrt{2}}\Bigl[\ud + \du\Bigr]$
																					& 1219 	& Recombine all ions to the same well\\
\Mgplus Doppler cool			& $\frac{1}{\sqrt{2}}\Bigl[\ud + \du\Bigr]$
																					& 400 	& -- \\
		--										& --						& 22 		& -- \\
$R(\pi,0)$ 								& $\ket{2,0}  \rightarrow \ket{2,-1}$
																					& 3			& Further transfer the residual spin populations $\epsilon_{A,B}$ \\
		--										&	--						& 22 		& -- \\
$R(\pi,0)$ 								& $\ket{2,-1} \rightarrow \ket{2,-2}$
																					& 4 		& Further transfer the residual spin populations $\epsilon_{A,B}$\\
		--										&	--						& 22 		& -- \\
$R^{c}(\frac{\pi}{2},-\frac{3\pi}{4})$
													& $\frac{1}{\sqrt{2}}\Bigl[\uu + i\dd\Bigr]$
																					& 1			& Rotate into the measurement basis \\
		--										&	--						& 6 		& -- \\
$R^{c}(\frac{\pi}{2},\phi_{p})$
													&	$\frac{1}{2}\Bigl[\left(\cos\phi_p-\sin\phi_p\right)
																							 \left(e^{-i\phi_p}\uu-e^{i\phi_p}\dd\right)
																		\linebreak	+ \left(\cos\phi_p+\sin\phi_p\right)\left(\ud+\du\right)\Bigr]$
														 							& 1 		& Analysis pulse with variable phase $\phi_{p}$\\
		--										&	--						& 79		& -- \\
$R(\pi,0)$ 								& $\ket{\down} \rightarrow \ket{2,0}$
																					& 3 		& Transfer the $\ket{\down}$ population \\
		--										&	--						& 22 		& -- \\
$R(\pi,0)$ 								& $\ket{2,0} \rightarrow \ket{2,-1}$
																					& 3 		& Further transfer the $\ket{\down}$ population \\
		--										&	--						& 22 		& -- \\
$R(\pi,0)$ 								& $\ket{2,-1}  \leftrightarrow \ket{2,-2}$
																					& 4 		& Further transfer the $\ket{\down}$ population
																												(and transfer any residual spin populations $\epsilon_{A,B}$ to $\ket{2,-1}$)\\
		--										&	--						& 43 		& -- \\
Measure $P_{\upup}, P_{\downdown},\newline P_{\updown}+P_{\downup}$
					 								& --						& 200 	& Determine the spin populations \\

\hline\hline

\textit{Total Time} 			&   						& $\sim$~14~ms & $\simeq$~600 laser pulses \\

\end{longtable*}

\end{scriptsize}

\end{document}